\documentclass[a4paper,11pt]{article}
\usepackage[dvips]{color,graphicx}
\usepackage{stmaryrd}
\def\dprod{\displaystyle\prod}

\def\dsum{\displaystyle\sum}

\usepackage{graphicx,amssymb,amsmath,bm,latexsym}

\allowdisplaybreaks

\begin{document}
\begin{center}

{\Large\bf  On $W_{1+\infty}$ $n$-algebra} \vskip .2in

{\large Chun-Hong Zhang$^{a}$,
Lu Ding$^{b}$,
Zhao-Wen Yan$^{c}$,\\
Ke Wu$^{a,d}$
and Wei-Zhong Zhao$^{a,d,}$
\footnote{Corresponding author: zhaowz@cnu.edu.cn}} \\
$^a${\em School of Mathematical Sciences, Capital Normal University,
Beijing 100048, China} \\
$^b${\em Institute of Applied Mathematics, Academy of Mathematics and Systems Science, Chinese
Academy of Sciences, Beijing 100190, China} \\
$^c${\em School of Mathematical Sciences, Inner Mongolia University, Hohhot
010021, China}\\
$^d${\em Beijing Center for Mathematics and Information
Interdisciplinary Sciences, Beijing 100048, China } \\

\begin{abstract}
We present the nontrivial $W_{1+\infty}$ $n$-algebra and analyze its remarkable properties.
We investigate the $W_{1+\infty}$ $n$-algebra in the Landau problem
and discuss the realization of the classical $w_{\infty}$ 3-algebra.
Furthermore, we discuss the case of the many-body system in the lowest Landau level
and derive the constraints for correlation functions of the vertex operators.

\end{abstract}

\end{center}

\section{Introduction}

Recently there has been a renewal of interest in $n$-algebra.
One found that there are the important applications of $n$-algebra
in the string theory \cite{BL2007, Gustavsson} and  condensed matter
physics \cite{Estienne}-\cite{Hasebe}.
The infinite-dimensional algebras have been vigorously investigated in the literature.
It is known that the Nambu 3-algebra can be regarded as a natural generalization of a
Lie algebra for higher-order algebraic operations \cite{Nambu, Takhtajan}.
A flurry of recentwork has been focused on the study of the infinite-dimensional Nambu 3-algebras,
such as Virasoro-Witt 3-algebra \cite{Curtright, Curtright2009},
(super) $w_{\infty}$ 3-algebra \cite{Chakrabortty, Chen2011} and $SDiff(T^3)$ 3-algebra \cite{Axenides}.
The $W_{\infty}$ algebra is the higher-spin extensions of the Virasoro algebra \cite{ShenPLB90, ShenNPB90}.
It naturally arises in various physical systems.
Its 3-algebra was derived in Ref.\cite{Chakrabortty} and the classical $w_{\infty}$ 3-algebra
was also presented. Although the Filippov condition or fundamental identity (FI) holds for
the classical $w_{\infty}$ 3-algebra, it fails for the $W_{\infty}$ 3-algebra.

The Landau problem is an interesting physical problem which deals with the motion of
a charged particle in a plane orthogonal to a uniform magnetic field.
The infinite-dimensional symmetry for the Landau problem has been well investigated.
One found that there exists the $W_{1+\infty}$ algebra in the free electron
theory of Landau levels \cite{Cappelli, Kogan}.
Moreover the $W_{1+\infty}$ algebra also
appears in  the quantum Hall effect \cite{Cappelli}-\cite{Karabali}.
It plays an important role for the incompressibility which is a property of
the bulk of the electron liquid. The $A$-class topological insulators can be
regarded as a higher dimensional counterpart of the quantum Hall effect.
Recently the $n$-algebraic structure of topological insulators has been
performed a detail study \cite{Estienne}.
It was found that there are the close relations between quantum Nambu bracket in even
dimensions and $A$-class topological insulator.
In this paper, we present the $W_{1+\infty}$ $n$-algebra and explore its remarkable
properties.  Then based on the well-known $W_{1+\infty}$ symmetry
in the Landau problem, we study its $n$-algebraic structures.

\section{ $W_{1+\infty}$ $n$-algebra}
Let us take the operators
\begin{eqnarray}\label{eq:walgoperator}
W_m^r=z^{m+r-1}(\frac{\partial}{\partial z})^{r-1}, \ \ r\in \mathbb{Z}_+, m \in\mathbb{N}.
\end{eqnarray}
We then obtain the $W_{1+\infty}$ algebra \cite{Cappelli}
\begin{eqnarray}\label{eq:walgebra}
[W_{m_1}^{r_1}, W_{m_2}^{r_2}]
=(\dsum_{\alpha=0}^{r_1-1}C_{r_1-1}^{\alpha}A_{m_2+r_2-1}^{\alpha}
-\dsum_{\alpha=0}^{r_2-1}C_{r_2-1}^{\alpha}
A_{m_1+r_1-1}^{\alpha})W_{m_1+m_2}^{r_1+r_2-1-\alpha},
\end{eqnarray}
where
$A_n^{\alpha}=\left\{\begin{array}{cc}
n(n-1)\cdots(n-\alpha+1),& \alpha\leqslant n,\\
0,                  &\alpha>n,\end{array}\right. $
and $C_n^{\alpha}=\frac{n(n-1)\cdots(n-\alpha+1)}{\alpha !}$.

Taking $r_1=r_2=2$ in (\ref{eq:walgebra}), it gives the Witt algebra
\begin{eqnarray}\label{VWalg}
[W_{m_1}^2, W_{m_2}^2]&=&(m_2-m_1)W_{m_1+m_2}^2.
 \end{eqnarray}

The $n$-commutator of the operators (\ref{eq:walgoperator}) yields
the $W_{1+\infty}$ $n$-algebra
\begin{eqnarray}\label{Wnalgebra}
[W_{m_1}^{r_1}, W_{m_2}^{r_2}, \ldots, W_{m_n}^{r_n}]&:=&\epsilon_{1 2 \cdots n}^{i_1 i_2 \cdots i_n}
W_{m_{i_1}}^{r_{i_1}}W_{m_{i_2}}^{r_{i_2}}\cdots W_{m_{i_n}}^{r_{i_n}}\nonumber\\
&=&\epsilon_{1 2 \cdots n}^{i_1 i_2 \cdots i_n} \dsum_{\alpha_1=0}^{\beta_1}
\dsum_{\alpha_2=0}^{\beta_2}
\cdots \dsum_{\alpha_{n-1}=0}^{\beta_{n-1}}
C_{\beta_1}^{\alpha_1}C_{\beta_2}^{\alpha_2}\cdots C_{\beta_{n-1}}^{\alpha_{n-1}}
\cdot A_{m_{i_2}+r_{i_2}-1}^{\alpha_1}\nonumber\\
&&A_{m_{i_3}+r_{i_3}-1}^{\alpha_2}\cdots A_{m_{i_n}+r_{i_n}-1}^{\alpha_{n-1}}
W_{m_1+m_2+\cdots +m_{n}}^{r_1+\cdots+r_n-(n-1)-\alpha_1-\cdots-\alpha_{n-1}},
\end{eqnarray}
where
$
\epsilon _{j_{1}\cdots j_{p}}^{i_{1}\cdots i_{p}}=\det \left(
\begin{array}{ccc}
\delta _{j_{1}}^{i_{1}} & \cdots & \delta _{j_{p}}^{i_{1}} \\
\vdots &  & \vdots \\
\delta _{j_{1}}^{i_{p}} & \cdots & \delta _{j_{p}}^{i_{p}}%
\end{array}%
\right)
$
and
$\beta_k=\left\{
  \begin{array}{cc}
    r_{i_1}-1,& k=1, \\
\dsum_{j=1}^k r_{i_j}-k-\dsum_{i=1}^{k-1}\alpha_i,   & 2\leqslant k\leqslant n-1.\\
  \end{array}\right.$

Due to the associative operators $W_m^r$ (\ref{eq:walgoperator}),
the $n$-algebra (\ref{Wnalgebra}) with $n$ even
satisfies the generalized Jacobi identity (GJI) \cite{Izquierdo}
\begin{equation}\label{ShJ}
\epsilon _{12\cdots 2n-1}^{i_{1}i_{2}\cdots i_{2n-1}}\left[ [B_{i_{1}},B_{i_{2}}\cdots
,B_{i_{n}}],B_{i_{n+1}},\cdots ,B_{i_{2n-1}}\right] =0.
\end{equation}
Hence the $W_{1+\infty}$ $n$-algebra (\ref{Wnalgebra}) with $n$ even is a
generalized Lie algebra  or higher order Lie algebra.
When $n$ is odd, the  GJI (\ref{ShJ}) does not hold for the $n$-algebra (\ref{Wnalgebra}).
In this case, it satisfies  the generalized Bremner identity (GBI) \cite{CJM09, Weingart}
\begin{eqnarray}\label{GBI}
&&\epsilon _{1\ldots 3n-3}^{i_{1}\ldots i_{3n-3}}
[[A,B_{i_{1}},\ldots ,B_{i_{n-1}}],[B_{i_{n}},\ldots
,B_{i_{2n-1}}],B_{i_{2n}},\ldots ,B_{i_{3n-3}}]  \notag\\
&&=\epsilon _{1\ldots 3n-3}^{i_{1}\ldots i_{3n-3}}
[[A,[B_{i_{1}},\ldots ,B_{i_{n}}],B_{i_{n+1}},\ldots
,B_{i_{2n-2}}],B_{i_{2n-1}},\ldots ,B_{i_{3n-3}}].
\end{eqnarray}

For the $W_{1+\infty}$ algebra (\ref{eq:walgebra}), it is well-known that there is only
a nontrivial subalgebra, i.e., the Witt algebra (\ref{VWalg}).
Therefore it is natural to pursue the subalgebra in (\ref{Wnalgebra}).
Let us consider the $2n$-commutator of the generators $ W_m^{n+1} $ with the fixed superindex $n+1\geqslant2$.
After a straightforward calculation we obtain the sub-$2n$-algebra
\begin{eqnarray}\label{sub2nalgebra}
[W_{m_1}^{n+1}, W_{m_2}^{n+1}, \ldots , W_{m_{2n}}^{n+1}]=\dprod_{1\leqslant j<k\leqslant 2n}(m_k-m_j)
W_{m_1+m_2+\cdots +m_{2n}}^{n+1},
\end{eqnarray}
where we take the scaled generators $W_{m}^{n+1}\rightarrow Q^{-\frac{1}{2n-1}}W_{m}^{n+1}$,
the scaling coefficient
$Q$ is given by
$Q=\dsum_{(\alpha_1,\alpha_2,\cdots,\alpha_{2n-1})\in S_{2n-1} }C_{\beta_1}^{\alpha_1}
C_{\beta_2}^{\alpha_2}\cdots  C_{\beta_{2n-1}}^{\alpha_{2n-1}}
\epsilon_{1 2 \cdots {(2n-1)}}^{\alpha_1 \alpha_2 \cdots \alpha_{2n-1}}$,
$S_{2n-1}$ is the permutation group of $\{1,2,\cdots,2n-1\}$.
When particularized to the $n=1$ case in (\ref{sub2nalgebra}), it gives
the Witt algebra (\ref{VWalg}).

Moreover, it is easy to obtain
\begin{eqnarray}\label{nullsubalg}
[W_{m_1}^{n+1}, \ldots ,W_{m_{2n+1}}^{n+1}]=0.
\end{eqnarray}

The centrally extended
$W_{1+\infty}$ algebra has been constructed in Refs.\cite{ShenPLB90,PRS}.
We see that the $W_{1+\infty}$ sub-$2n$-algebra is a generalized Lie algebra
and, further, its structure constants  are determined by the Vandermonde determinant.
Owing to remarkable property of the structure constants,
we may derive the centrally extended $W_{1+\infty}$ sub-$2n$-algebra
\begin{eqnarray}\label{csubW2nalgebra}
&& [W_{m_1}^{n+1}, W_{m_2}^{n+1}, \ldots , W_{m_{2n}}^{n+1}]=\dprod_{1\leqslant j<k\leqslant 2n}(m_k-m_j)
W_{m_1+m_2+\cdots +m_{2n}}^{n+1}\nonumber\\
&&+\frac{c}{12 \times2^n n!}\dsum_{(i_1i_2\cdots i_{2n}) \in S_{2n}}
\epsilon_{1 2 \cdots {2n}}^{i_1 i_2 \cdots i_{2n}}
\dprod_{k=1}^n[(m_{i_{2k-1}}^3-m_{i_{2k-1}})\delta_{m_{i_{2k-1}}+m_{i_{2k}},0}],
\end{eqnarray}
where $c$ is an arbitrary constant.

As the examples, taking the generators $W_{m}^2$ and $W_{m}^3$ in (\ref{csubW2nalgebra}), respectively,
we have the Virasoro algebra
\begin{eqnarray}
[W_{m_1}^2,W_{m_2}^2]
&=&(m_2-m_1)W_{m_1+m_2}^2+\frac{c}{12}(m_1^3-m_1)\delta_{m_1+m_2,0},
 \end{eqnarray}
and centrally extended $4$-algebra
\begin{eqnarray}
[W_{m_1}^3, W_{m_2}^3, W_{m_3}^3, W_{m_4}^3]
&=&(m_4-m_3)(m_4-m_2)(m_4-m_1)(m_3-m_2)\nonumber\\
&&(m_3-m_1)(m_2-m_1)W_{m_1+m_2+m_3+m_4}^3\nonumber\\
&+&\frac{c}{12}[(m_1^3-m_1)(m_3^3-m_3)\delta_{m_1+m_2,0}\delta_{m_3+m_4,0}\nonumber\\
&+&(m_2^3-m_2)(m_3^3-m_3)\delta_{m_1+m_3,0}\delta_{m_2+m_4,0}\nonumber\\
&+&(m_1^3-m_1)(m_2^3-m_2)\delta_{m_1+m_4,0}\delta_{m_2+m_3,0}].
\end{eqnarray}

\section{$W_{1+\infty}$ $n$-algebra in the Landau problem}
Let us consider an electron of charge $e$ and mass $m$ moving on a plane in
presence of a constant perpendicular magnetic field $B$.
The Hamiltonian is given by
\begin{eqnarray}\label{Hamiltonian}
H=\frac{1}{2m}(\mathbf{p}-\frac{e}{c}\mathbf{A})^2,
\end{eqnarray}
where the momentum $\mathbf{p}=-\mathbf{i}\hbar \mathbf{\nabla}$ and
the gauge potential $\mathbf{A}$ exist in the plane.
Let us choose the symmetric $\mathbf{A}=\frac{B}{2}(-y,x)$ and introduce complex
variables $z=x+\mathbf{i} y ,\bar{z}=x-\mathbf{i} y $.
In the following the units $c=m=1$ will be employed. For convenience we take $eB=2$.

We may construct the harmonic oscillator operators $a$ and $a^\dag$,
\begin{eqnarray}
&&a=\frac{z}{2}+ \hbar{\bar{\partial}}, \ \ a^\dag=\frac{\bar{z}}{2}-\hbar{\partial},
\end{eqnarray}
obeying
\begin{eqnarray}
[a, a^\dag]=\hbar.
\end{eqnarray}
In terms of the harmonic oscillator operators $a$ and $a^\dag$,
the Hamiltonian (\ref{Hamiltonian}) can be rewritten as
\begin{eqnarray}\label{Hamiltonian1}
&&H=a a^{\dag}+a^{\dag}a.
\end{eqnarray}

We may introduce another pair  $b$ and $b^\dag$ commuting with  $a$ and $a^\dag$,
\begin{eqnarray}
&&b=\frac{\bar{z}}{2}+\hbar {\partial}, \ \ b^\dag=\frac{z}{2}-\hbar\bar{\partial},
\end{eqnarray}
with commutation relation
\begin{eqnarray}\label{eq:bcomm}
[b, b^\dag]=\hbar.
\end{eqnarray}
The angular momentum operator can be written as
\begin{eqnarray}\label{amo}
&&J= b^{\dag}b-a^{\dag}a.
\end{eqnarray}
It is commuting with the Hamiltonian (\ref{Hamiltonian1}), i.e.,
\begin{eqnarray}\label{CJH}
[J, H]=0.
\end{eqnarray}

Let us take the operators \cite{Cappelli}
\begin{eqnarray}\label{eq:woperator1}
{\tilde W}_m^r=( b^\dag)^{m+r-1}(b)^{r-1},\  r \geqslant 1, \  m+r \geqslant 1,
\end{eqnarray}
which are commuting with the Hamiltonian  (\ref{Hamiltonian1}),
\begin{eqnarray}\label{CWH}
[\tilde W_{m}^{r}, H]=0.
\end{eqnarray}

For the infinite conserved operators (\ref{eq:woperator1}), they form
the $W_{1+\infty}$ algebra \cite{Cappelli}
\begin{eqnarray}\label{Lwalgebra}
[\tilde W_{m_1}^{r_1}, \tilde W_{m_2}^{r_2}]
=(\dsum_{\alpha=0}^{r_1-1}C_{r_1-1}^{\alpha}A_{m_2+r_2-1}^{\alpha}
-\dsum_{\alpha=0}^{r_2-1}C_{r_2-1}^{\alpha}
A_{m_1+r_1-1}^{\alpha})\hbar^\alpha \tilde W_{m_1+m_2}^{r_1+r_2-1-\alpha},
\end{eqnarray}
and $n$-algebra
\begin{eqnarray}\label{LWnalgebra}
[\tilde W_{m_1}^{r_1}, \tilde W_{m_2}^{r_2}, \ldots, \tilde W_{m_n}^{r_n}]
&=&\epsilon_{1 2 \cdots n}^{i_1 i_2 \cdots i_n} \dsum_{\alpha_1=0}^{\beta_1}
\dsum_{\alpha_2=0}^{\beta_2}
\cdots \dsum_{\alpha_{n-1}=0}^{\beta_{n-1}}
C_{\beta_1}^{\alpha_1}C_{\beta_2}^{\alpha_2}\cdots C_{\beta_{n-1}}^{\alpha_{n-1}}
\cdot A_{m_{i_2}+r_{i_2}-1}^{\alpha_1}A_{m_{i_3}+r_{i_3}-1}^{\alpha_2}\nonumber\\
&&\cdots A_{m_{i_n}+r_{i_n}-1}^{\alpha_{n-1}}
\hbar^{\alpha_1+\cdots+\alpha_{n-1}}\tilde W_{m_1+m_2+\cdots +m_{n}}^{r_1+\cdots+r_n-(n-1)-\alpha_1-\cdots-\alpha_{n-1}}.
\end{eqnarray}
It should be pointed out that not as the case of the generators (\ref{eq:walgoperator}),
here the ($n$-)algebra corresponds to the so-called ¡°wedge¡± $W_{\Lambda}=\{\tilde W_m^r, |m|\leqslant r-1\}$,
plus the positive modes $m>r-1$.

In the classical limit, (\ref{Lwalgebra}) reduces to the classical $w_{\infty}$ algebra \cite{ShenPLB90,Cappelli}
\begin{eqnarray}\label{classicalW}
\{\tilde W_{m_1}^{r_1}, \tilde W_{m_2}^{r_2}\}=-\mathbf{i}(m_2(r_1-1)-m_1(r_2-1))\tilde W_{m_1+m_2}^{r_1+r_2-2},
\end{eqnarray}
which is the canonical symmetry of the classical case of the Hamiltonian (\ref{Hamiltonian}).

From (\ref{LWnalgebra}), we have the $W_{1+\infty}$ $3$-algebra
\begin{eqnarray}\label{rW3alg}
[\tilde W_{m_1}^{r_1}, \tilde W_{m_2}^{r_2}, \tilde W_{m_3}^{r_3}]
=\hbar(r_1(m_2-m_3)+r_2(m_3-m_1)
+r_3(m_1-m_2))\tilde W_{m_1+m_2+m_3}^{r_1+r_2+r_3-3}+O(\hbar^2).
\end{eqnarray}
It satisfies the Bremner identity (BI) \cite{Bremner1998, Bremner2006}
\begin{eqnarray}\label{eq:BI}
\epsilon^{i_1i_2\cdots i_6}[[A, [B_{i_1}, B_{i_2}, B_{i_3}], B_{i_4}], B_{i_5}, B_{i_6}]
=\epsilon^{i_1i_2\cdots i_6}[[A, B_{i_1}, B_{i_2}], [B_{i_3}, B_{i_4}, B_{i_5}], B_{i_6}] ,
\end{eqnarray}
where $i_1, \cdots, i_6$ are implicitly summed from 1 to 6.

In the classical limit
$\{\ , \ ,\}=\lim\limits_{\hbar \rightarrow 0 }\frac{1}{{\bf
i}\hbar}[\ ,\ , ]$,
(\ref{rW3alg}) becomes the classical $w_{\infty}$-3-algebra \cite{Chakrabortty}
\begin{eqnarray}\label{lw3alg}
\{\tilde W_{m_1}^{r_1}, \tilde W_{m_2}^{r_2}, \tilde W_{m_3}^{r_3}\}=
-\mathbf{i} (r_1(m_2-m_3)+r_2(m_3-m_1)+r_3(m_1-m_2))\tilde W_{m_1+m_2+m_3}^{r_1+r_2+r_3-3}.
\end{eqnarray}
It is worth to emphasize that (\ref{lw3alg})
satisfies the classical FI \cite{Filippov}
\begin{eqnarray}\label{eq:FI}
\{A, B, \{C, D, E\}\}=\{\{A, B, C\}, D, E\}+\{C, \{A, B, D\}, E\}+\{C, D, \{A, B, E\}\}.
\end{eqnarray}

Let us pause here to look at the classical $w_{\infty}$ algebra (\ref{classicalW})
and its 3-algebra (\ref{lw3alg}).
For the classical case of the Hamiltonian (\ref{Hamiltonian}), the most general canonical transformations
which leave it invariant are generated by  $\tilde W_{m,r}^{cl}=(b^\dag)^{m+r-1}b^{r-1}$,
where $b=\frac{1}{2}((p_y+x)+\mathbf{i} (p_x-y))$ and $b^{\dag}=\frac{1}{2}((p_y+x)-\mathbf{i} (p_x-y))$.
Cappelli {\it et al.} \cite{Cappelli} showed that
$\tilde W_{m,r}^{cl}$ act nontrivially only on a two-dimensional subspace
which admits a symplectic structure in terms of $b$ and $b^{\dag}$,
and the usual Poisson bracket $\{\tilde W_{m_1,r_1}^{cl}, \tilde W_{m_2,r_2}^{cl}\}$
gives the classical $w_{\infty}$ algebra (\ref{classicalW}).

The classical Nambu 3-bracket \cite{Nambu} is defined for a triple of
classical observables on the three-dimensional phase space
$\mathbb{R}^3$ with coordinates $x, y, z$ by the formula
$\{f, g, h\}=\frac{\partial (f, g, h)}{\partial (x,y,z)}$.
Based on the classical Nambu 3-bracket,
Chakrabortty ${\it et \ al}.$ \cite{Chakrabortty} presented the realization
of the classical $w_{\infty}$ 3-algebra (\ref{lw3alg}).
However, it is noted that the above mentioned Poisson bracket
$\{\tilde W_{m_1,r_1}^{cl}, \tilde W_{m_2,r_2}^{cl}\}$ gives
the $w_{\infty}$ algebra of area-preserving diffeomorphisms of the two-dimensional plane.
It is obvious that the classical Nambu 3-bracket  can not be used here.
Not as done in Ref.\cite{Chakrabortty}, let us consider the Dzhumadildaev 3-bracket \cite{Dzhumadildaev, Kac}
\begin{eqnarray}\label{3wronsikian}
\{f,g,h\}&=&\left|
              \begin{array}{ccc}
                f & g & h \\
                \frac{\partial f}{\partial x} & \frac{\partial g}{\partial x}  & \frac{\partial h}{\partial x}  \\
                \frac{\partial f}{\partial y} & \frac{\partial g}{\partial y}  & \frac{\partial h}{\partial y} \\
              \end{array}
            \right|.
\end{eqnarray}
It should be pointed out that the Dzhumadildaev $3$-bracket
satisfies the skew-symmetry and FI (\ref{eq:FI}), but the Leibniz rule does not hold for it.
The alternative is the following generalized Leibniz rule:
\begin{eqnarray}
\{f_1, f_{2}, gh\}=\{f_1, f_{2}, g\}h+g\{f_1, f_{2}, h\}-\{f_1, f_{2}, 1\}gh.
\end{eqnarray}
In terms of the Dzhumadildaev $3$-bracket, it is easy to verify that the generators $\tilde W_{m,r}^{cl}$
yield the $w_{\infty}$ 3-algebra (\ref{lw3alg}) which is not a Nambu 3-algebra
but the so-called Dzhumadildaev $3$-algebra.

For the infinite conserved operators (\ref{eq:woperator1}), they also form
the sub-$2n$-algebra
\begin{eqnarray}\label{sub2nalgebrahbar}
[\tilde W_{m_1}^{n+1}, \tilde W_{m_2}^{n+1}, \ldots , \tilde W_{m_{2n}}^{n+1}]
=\hbar^{n(2n-1)} \dprod_{1\leqslant j<k\leqslant 2n}(m_k-m_j)
\tilde W_{m_1+m_2+\cdots +m_{2n}}^{n+1}.
\end{eqnarray}
Note that (\ref{sub2nalgebrahbar})
contains only a unique term in $\hbar^{n(2n-1)}$.
It represents the intrinsic symmetry in higher order of $\hbar$.

For the $2n$-commutator, there is the limiting relation \cite{Zachos2003}
\begin{eqnarray}\label{limnb}
\frac{1}{n!}\lim\limits_{\hbar \rightarrow 0 }(\frac{1}{{\bf
i}\hbar})^n[B_1 , B_2, \cdots, B_{2n}]=\{B_1, B_2,\cdots, B_{2n}\},
\end{eqnarray}
where the right hand side bracket is the classical Nambu $2n$-bracket.
In the classical limit (\ref{limnb}), the $W_{1+\infty}$ sub-$2n$-algebra
(\ref{sub2nalgebrahbar}) becomes
\begin{eqnarray}\label{nsub2nalg}
\{\tilde W_{m_1}^{n+1}, \tilde W_{m_2}^{n+1}, \ldots , \tilde W_{m_{2n}}^{n+1}\}=0.
\end{eqnarray}

It is known that the angular momentum operator $J$ (\ref{amo}) and
the operators (\ref{eq:woperator1}) are
commuting with the Hamiltonian $H$ (\ref{Hamiltonian1}), respectively.
For the triple operators $(\tilde W_m^r, J, H)$, we note that when $m=0$,
they are in involution with the 3-bracket structure, i.e.,
\begin{eqnarray}\label{w0jh}
[\tilde W_0^r, J, H]=0.
\end{eqnarray}
Using (\ref{sub2nalgebrahbar}),
(\ref{w0jh}) can be rewritten as
\begin{eqnarray}\label{w0rjh}
[[\tilde W_{m_1}^{r}, \tilde W_{m_2}^{r}, \ldots , \tilde W_{m_{2r-2}}^{r}], J, H]=0,
\end{eqnarray}
where $m_1+m_2+\cdots+m_{2r-2}=0$.

The one-body state of an electron in the lowest Landau level is given by the
Landau site $|n\rangle$. In the symmetric gauge it is the angular-momentum
state $|n\rangle$ defined by
\begin{eqnarray}
|n \rangle=\frac{1}{\sqrt{n!}}(b^\dag)^n|0 \rangle,\ \ n=0,1,2,\cdots.
\end{eqnarray}

In terms of the operators (\ref{eq:woperator1}), two of the states are related as
\begin{eqnarray}\label{staterel}
|m+n \rangle=\frac{1}{\hbar^n\sqrt{n!(m+n)!}}\tilde W_{m}^{n+1}|n \rangle.
\end{eqnarray}
Since there is the sub-$2n$-algebra (\ref{sub2nalgebrahbar}),
(\ref{staterel}) can be expressed as
\begin{eqnarray}\label{staterel2}
|m+n \rangle \thicksim\tilde W_m^{n+1}|n \rangle
\thicksim [\tilde W_{m_1}^{n+1},  \ldots , \tilde W_{m_{2n}}^{n+1}]|n \rangle,
\end{eqnarray}
where $m=\sum_{i=1}^{2n}m_{i}$, the coefficients are omitted and
$\thicksim$ means that these two states are equivalent.
It is of interest to note that for the given $m$ and $n$,
the sub-$2n$-algebra in (\ref{staterel2}) indicates the refined relation between
the two states $|m+n \rangle$ and $|n \rangle$.

It should be noted that the operators $\tilde W_m^r$ generically
involve powers of derivatives higher than one, and
therefore are not generators of local coordinate transformations on the wave
function. It implies that the generators $\tilde W_m^r$
are quasi-local operators. The generating function of $\tilde W_m^r$ is
actually the finite magnetic translation.

Let us take \cite{Kogan}
\begin{eqnarray}\label{eq:woperator}
T_{\bm n}=exp(-\frac{B}{2}|\bm n|^2)\sum^{\infty}_{k,l=0}(-1)^l\frac{(n_1+\textbf{i} n_2)^k}{k!}
\frac{(n_1-\textbf{i} n_2)^l}{l!}\tilde W_{k-l}^{l+1},
\end{eqnarray}
where ${\bm n}=(n_1,n_2)$, $b=\frac{B\bar{z}}{2}+ \partial$, $b^\dag=\frac{B z}{2}- \bar{\partial}$ and
the commutator between $b$ and $b^\dag$ is $[b, b^\dag]=B$.

The operators (\ref{eq:woperator}) yield the sine algebra \cite{Fairlie224, Fairlie218}
\begin{eqnarray}\label{eq:sinealg}
[T_{\bm{n}}, T_{\bm{m}}]
&=&2{\bf i}\sin B(\bm{m}\times \bm{n})T_{\bm{n+m}},
\end{eqnarray}
and (co)sine $n$-algebra \cite{Ding}
\begin{eqnarray}\label{sinnalg}
&& [T_{\bm{m_1}},\cdots ,T_{\bm{m_n}}]
=\frac{1}{2}( \epsilon_{1\cdots n}^{i_{1}\cdots i_{n}}\exp
( \mathbf{i}B\sum_{l<k}\bm{m_{i_k}}\times\bm{m_{i_l}}) \notag \\
&&+\epsilon_{1\cdots n}^{i_{n}\cdots i_{1}}\exp (\mathbf{i} B\sum_{l>k}\bm{m_{i_k}}
\times\bm {m_{i_l}}) ) T_{\bm {m_1+\cdots+m_n}}\notag \\
&&=\frac{1}{2}\epsilon_{1\cdots n}^{i_{1}\cdots i_{n}}
\cos (B\sum_{l<k}\bm{m_{i_k}}\times\bm{m_{i_l}}) ( 1+( -1) ^{%
\frac{n( n-1) }{2}}) T_{\bm{m_1+\cdots+m_n}}  \notag \\
&&+\frac{\mathbf{i}}{2}\epsilon_{1\cdots n}^{i_{1}\cdots i_{n}}\sin
( B\sum_{l<k}\bm{m_{i_k}}\times\bm{m_{i_l}}) ( 1-(
-1) ^{\frac{n( n-1) }{2}}) T_{\bm{m_1+\cdots+m_n}}.
\end{eqnarray}
When $n$ is even, (\ref{sinnalg}) is a generalized Lie algebra.

Corresponding to (\ref{eq:FI}), the FI for the quantal ternary algebras is
\begin{eqnarray}\label{eq:qFI}
[A, B, [C, D, E]]=[[A, B, C], D, E]+[C, [A, B, D], E]+[C, D, [A, B, E]].
\end{eqnarray}
Note that it is not an operator identity, which holds only in special circumstances.
It is easy to verify that (\ref{eq:qFI}) does not hold for
the following sine 3-algebra with the general parameter $B$:
\begin{eqnarray}
\lbrack T_{\bm {m}},T_{\bm{n}},T_{\bm{k}}]
&=&2\mathbf{i}[\sin B(\bm{m}\times \bm{n}-\bm{n}\times\bm{k}
-\bm{k}\times\bm{m})  \notag  \label{Psin3} \\
&&+\sin B(\bm{k}\times\bm{m}-\bm{m}\times\bm{n}
-\bm{n}\times\bm{k})  \notag \\
&&+\sin B(\bm{n}\times\bm{k}-\bm{k}\times\bm{m}
-\bm{m}\times\bm{n})]T_{\bm{m+n+k}}.
\end{eqnarray}
However, when particularized to the $B =\frac{\pi}{2}$ case,
(\ref{Psin3}) becomes a Filippov algebra  which satisfies the FI (\ref{eq:qFI}) \cite{Ding}.

For the arbitrary $B$, (\ref{sinnalg}) with $n$ odd is also not a generalized Lie algebra. However,
Ding ${\it et \ al}.$ \cite{Ding} found that when $B =\frac{\pi}{3}$,
the cosine 5-algebra (\ref{sinnalg}) becomes a generalized Lie algebra.
For the general case of (\ref{sinnalg}), let us present a conjecture as follows:\\
{\it  When $B =\frac{\pi}{n+1}$, the (co)sine $(2n+1)$-algebra (\ref{sinnalg})
is a generalized Lie algebra.}

\section{The many-body system in the lowest Landau level}
In the previous section, we have investigated the $W_{1+\infty}$ $n$-algebra
of the infinite conserved operators for an electron  moving on a plane in
presence of a constant perpendicular magnetic field.
Let us turn to the case of $N$ electrons \cite{Cappelli}.
The Hamiltonian is given by
\begin{eqnarray}\label{Hamiltonian2}
\bar H=\frac{1}{2m}\dsum_{k=1}^{N}[\mathbf{p}_k-\frac{e}{c}\mathbf{A}(\mathbf{r}_k)]^2.
\end{eqnarray}

We may construct the harmonic oscillator operators $a$ and $a^\dag$,
\begin{eqnarray}\label{oscillator}
&&a_k=\frac{z_k}{2}+ \hbar{\bar \partial}_k, \ \ a_k^\dag=\frac{\bar z_k}{2}-\hbar{\partial_k},
\end{eqnarray}
obeying
\begin{eqnarray}
[a_k, a_l^\dag]=\hbar\delta_{kl},
\end{eqnarray}
where $z_k=x_k+{\bf i}y_k$ is the complex coordinate for the location of the $k$th electron.

In the appropriately chosen system of units $c=m=1$ , $eB=2$
and symmetric gauge $\mathbf{A}=\frac{B}{2}(-y,x)$,
in terms of harmonic oscillator operators (\ref{oscillator}),
the Hamiltonian (\ref{Hamiltonian2}) can be rewritten as
\begin{eqnarray}
&&\bar H=\dsum_{k=1}^{N}(a_k a_k^{\dag}+a_k^{\dag}a_k).
\end{eqnarray}

We may also express the angular momentum of $N$ electrons as
\begin{eqnarray}
\bar J=\hbar\dsum_{k=1}^{N}(z_k\partial_k-{\bar z}_k{\bar \partial}_k)
=\dsum_{k=1}^{N}[b_k^{\dag}b_k-a_k^{\dag}a_k],
\end{eqnarray}
where
\begin{eqnarray}
&&b_k=\frac{{\bar z}_k}{2}+\hbar \partial_k, \ \ b_k^\dag=\frac{z_k}{2}-\hbar{{\bar \partial}_k},
\end{eqnarray}
obeying
\begin{eqnarray}
[b_k, b_l^\dag]=\hbar\delta_{kl}.
\end{eqnarray}

Let us take the operators \cite{Cappelli}
\begin{eqnarray}\label{Nwalgebra}
{\bar W}_m^r=\dsum_{i=1}^N(b_i^\dag)^{m+r-1}(b_i)^{r-1},
\end{eqnarray}
which are commuting with the Hamiltonian  (\ref{Hamiltonian2}).

The infinite conserved operators (\ref{Nwalgebra}) also form
the $W_{1+\infty}$ algebra.
However, they do not yield the nontrivial $n$-algebra
except for
\begin{eqnarray}\label{conclusion}
[\bar W_{m_1}^{n+1}, \bar W_{m_2}^{n+1}, \cdots, \bar W_{m_{2nN+1}}^{n+1}]=0.
\end{eqnarray}
Here we should like to draw attention to the fact that the symmetry (\ref{conclusion})
is different from (\ref{nullsubalg}). It is not only determined by the superindex of the generators,
but also the number of electrons $N$.
For the triple operators $(\bar W_m^r, \bar J, \bar H)$, we also have
(\ref{w0jh}). However, not as the case of single electron, (\ref{w0rjh}) does not hold.

It is well-known that the filling factor $\nu=1$ ground state is given by the wave function
\begin{eqnarray}\label{wavefunction}
\psi_1(z_1,\cdots, z_N)=\dprod_{1\leqslant i<j\leqslant N}(z_i-z_j)exp(-\frac{1}{2\hbar}\dsum_{i=1}^N|z_i|^2).
\end{eqnarray}
In the limit of weak coupling, Fubini \cite{Fubini} introduced the state of minimum angular momentum which
is given by the Vandermonde determinant
\begin{eqnarray}\label{smam}
f(z_1,\cdots, z_N)=\dprod_{1\leqslant i<j\leqslant N}(z_i-z_j).
\end{eqnarray}

Let us take the operators
\begin{eqnarray}\label{mulwalgoperator1}
\hat W_m^r&=&(\frac{1}{2\hbar})^{r-1}\sum\limits_{i=1}^N(b_i^\dag+a_i)^{m+r-1}
(b_i-a_i^\dag)^{r-1}\nonumber\\
&=&\sum\limits_{i=1}^Nz_i^{m+r-1}(\frac{\partial}{\partial z_i})^{r-1},
\ \ r\in \mathbb{Z}_+, m \in\mathbb{N},
\end{eqnarray}
which yield the $W_{1+\infty}$ algebra and null $(2nN+1)$-algebra (\ref{conclusion})
as the case of (\ref{Nwalgebra}).

Moreover, when $N=n+1$, we have
\begin{eqnarray}
[ \hat W_{m_1}^{n+1}, \cdots, \hat W_{m_{2n}}^{n+1}]f(z_1,\cdots, z_N)
&=&\dprod_{1\leqslant i<j\leqslant 2n}(m_i-m_j) \hat W_{\bar m}^{n+1}f(z_1,\cdots, z_N)
\end{eqnarray}
and
\begin{eqnarray}\label{swer}
&&[ \hat W_{m_1}^{n+1}, \cdots, \hat W_{m_{2n+1}}^{n+1}]f(z_1,\cdots, z_N)=0,
\end{eqnarray}
where we take the scaled generators as done in (\ref{sub2nalgebra}).
We see that  the quantum multibrackets for the generators with multi-variable realization
(\ref{mulwalgoperator1})
are compatible with the results of (\ref{sub2nalgebra}) and (\ref{nullsubalg}) once we act on the
state of minimum angular momentum (\ref{smam}).

Let us introduce the Fubini-Veneziano vertex operator
\begin{eqnarray}\label{VO}
V(z)=:e^{{\bf{i}}Q(z)}:\equiv e^{{\bf{i}}Q^{+}(z)}e^{{\bf{i}}q}e^{-p\ln z}e^{{\bf{i}}Q^{-}(z)},
\end{eqnarray}
where $: :$ symbol denotes normal order,
$Q(z)=q-{\bf{i}}p\ln z+Q^{+}(z)+Q^{-}(z)$, $Q^{+}(z)=-{\bf{i}}\sum a_n^{\dag}z^n$,
$Q^{-}(z)={\bf{i}}\sum a_nz^{-n}$,
the zero mode operators $q$ and $p$ satisfy the commutation relation
$[q, p]=\bf{i}$, the infinite set of creation and destruction operators
$a_n$ and $a_n^\dag$ satisfy $[a_n, a_m^\dag]=\frac{1}{n}\delta_{nm}$.

The vertex operator (\ref{VO}) can be used to reproduce (\ref{smam})
as correlation functions \cite{Fubini}, i.e.,
\begin{eqnarray}\label{rcf}
<N|V(z_1)\cdots V(z_N)|0>=\dprod_{1\leqslant i<j\leqslant N}(z_i-z_j),
\end{eqnarray}
where $|0>$ and $<N|$ are the Fock  and charged vacuums, respectively.

From (\ref{swer}) and (\ref{rcf}), we obtain the constraints
for correlation functions
\begin{eqnarray}
[ \hat W_{m_1}^{n+1}, \cdots, \hat W_{m_{2n+1}}^{n+1}]
<n+1|V(z_1)\cdots V(z_{n+1})|0>=0.
\end{eqnarray}

\section{Conclusions}

We have derived the nontrivial $W_{1+\infty}$ $n$-algebra
which is the so-called generalized Lie algebra for the $n$ even case.
It is known that the Witt algebra is the unique nontrivial subalgebra
in $W_{1+\infty}$ algebra. The remaining generators do not yield any
nontrivial subalgebra. Our key finding is that the generators
(\ref{eq:walgoperator}) with any fixed superindex yield the nontrivial
sub-$2n$-algebra in which the structure constants are determined by
the Vandermonde determinant. Due to the remarkable property of the
structure constants, we also derived the centrally extended $W_{1+\infty}$
sub-$2n$-algebra. Since the form of $W_{1+\infty}$ $2n$-algebra
appears to become more complicated, it still remains on open question whether
there exist the central extension terms for the general $W_{1+\infty}$ $2n$-algebra.

It is known that there exist infinite conserved operators in the Landau problem,
which yield the $W_{1+\infty}$ algebra. Based on the quantum multibrackets,
we found that these infinite conserved operators also yield the $W_{1+\infty}$ $n$-algebra.
A remarkable property is that in the classical limit, these infinite-dimensional symmetries
do not only reduce to the well-known classical $w_{\infty}$ algebra, but also give the
$w_{\infty}$ 3-algebra which is the so-called Dzhumadildaev $3$-algebra.
Moreover, we have studied the case of the many-body problem in the lowest Landau level.
Since the closed $W_{1+\infty}$ sub-$2n$-algebra arises once we act on the
state of minimum angular momentum, we have derived the constraints for correlation functions
of the vertex operators. Our analysis provides additional
insight into the $W_{1+\infty}$ $n$-algebra. Due to its remarkable properties,
more applications in physics should be of interest.

\section*{Acknowledgement}
This work is supported by the National Natural Science Foundation of China (Nos. 11875194,
11375119, 11475116 and 11605096).


\end{document}